\documentclass{fizik}%
\usepackage{multicol,epsfig}

\vol{27}
\pyear{2003}
\received{\today}
\author[Entin-Wohlman, Aharony, Kashcheyevs]{
\textbf{Ora ENTIN-WOHLMAN, Amnon AHARONY, and Vyacheslavs KASHCHEYEVS}\\
\textit{School of Physics and Astronomy, Raymond and Beverly
Sackler
Faculty of Exact Sciences,}\\
\textit{Tel Aviv University, Tel Aviv 69978,
ISRAEL}\\
}

\title{Adiabatic Quantum Pumping of Coherent Electrons}

\begin{document}
\maketitle

\begin{abstract}
We review recent theoretical calculations of charge transfer
through mesoscopic devices in response to slowly-oscillating,
spatially-confined, potentials. The discussion is restricted to
non-interacting electrons, and emphasizes the role of quantum
interference and resonant transmission in producing almost integer
values (in units of the electronic charge $e$) of the charge
transmitted per cycle, $Q$. The expression for the pumped charge
is derived from a systematic expansion of the system scattering
states in terms of the temporal derivatives of the instantaneous
solutions. This yields the effect of the modulating potential on
the Landauer formula for the conductance in response to a constant
bias on one hand, and the corrections to the widely-used
adiabatic-limit formula (in which the modulation frequency is
smaller than any electronic relaxation rate) on the other hand.
The expression for $Q$ is used in connection with simple models to
exemplify the intimate relationship between resonant transmission
through the mesoscopic device and almost integral values of $Q$,
and to analyze the charge pumped by a surface acoustic wave
coupled to a quantum channel by the piezoelectric effect.
\end{abstract}

\section{Introduction}

Although electric charge consists of units of the electronic
charge $e$, the electric current in macroscopic samples is not
quantized, but rather behaves as a continuous fluid. The ability
to control current on the level of single electrons is feasible in
mesoscopic devices, and at low enough temperatures. The
theoretical possibility of current quantization has been first
addressed by Thouless \cite{thouless}, who considered the direct
current (dc) induced in a one-dimensional gas of non-interacting
electrons by a slowly-moving periodic potential profile. Thouless
has shown that when the Fermi energy lies in the gap between the
filled and empty bands of the time-dependent Hamiltonian (and the
gap does not close within the cycle), the current transmitted
through a cross-section of the sample over a single period $T$ of
the potential corresponds to an integral number of $e$:
$I=(e/T)\times$integer, where $T=a/v$, with $a$ and $v$ being the
lattice constant of the potential profile and its velocity. The
robustness of the quantization in such systems, with respect to
the effect of disorder or of many-body interactions, has been
further discussed in Refs. \cite{niu}. The charge transmitted in
this way is independent of the frequency of the modulating
potential (and hence the current per period is proportional to the
frequency).

The model of Thouless is a remarkably clear example of the
phenomenon termed `charge pumping' \cite{hekking,altshuler}.
Experimentally, investigations of this phenomenon focus on
confined nanostructures, e.g., quantum dots, carbon nanotubes, or
quantum channels. In the latter, traveling potential profiles,
like the one required in the Thouless model, can be generated
using the piezo-electric coupling of surface acoustic waves to the
two-dimensional electron gas formed in GaAs-AlGaAs
heterostructures \cite{Sh96}. Another possibility is to modulate
periodically gate voltages applied to quantum-dot devices
\cite{geerligs,Kou91,Po92,haz}. Exploiting the Coulomb-blockade
effect, this leads to quantization of the transmitted charge, and
is of metrological use \cite{co}. Devices of this type are called
`single-electron turnstiles'.

Charge pumping is achieved by varying certain parameters of the
system Hamiltonian periodically with time. When the change is
carried out slowly enough, the Hamiltonian returns to itself after
each cycle, and adiabatic control of the electronic states becomes
feasible. Under these circumstances, the modulating potential
affects predominantly the quantum interference of the electrons in
the structure, not their classical trajectories. This situation is
called `adiabatic quantum pumping'
\cite{spi,b,z,s,simon,a,an,a1,w,lew,mi,va,do}, and necessitates
the maintenance of  a phase-coherent motion of the electrons
(effects of dissipation, inelastic scattering, noise and dephasing
on quantum pumping have been discussed in Refs. \cite{m,va1,cr}).
Quantum pumping has been realized in an `open' quantum dot under
conditions in which Coulomb-blockade effects are not important
\cite{Sw99}.  The device of Ref. \cite{Sw99} consists of a
semi-conductor quantum dot, coupled to electronic reservoirs by
ballistic point-contacts. By varying gate voltages, the shape of
the confining potential of the electrons in the dot has been
modulated periodically, leading to a dc potential difference as a
consequence of electronic transfer between reservoirs. The
importance of quantum interference in determining the magnitude
and direction of the pumped charge has been clearly demonstrated
by applying a weak magnetic field perpendicular to the sample
\cite{Sw99}. Recently, the possibility of spin-polarized pumped
currents has been discussed \cite{tom,mu}.

Whereas a highly-accurate quantized pumped current has been
observed in the turnstile device \cite{Kou91} (in which the
quantization is dictated by the Coulomb-blockade effect), this is
not the case in the open-dot geometry \cite{Sw99} (where this
effect is expected to play a minor role). It is therefore of
fundamental interest to study charge pumping of non-interacting
electrons resulting from interference effects, to explore the
circumstances under which it is optimal \cite{a,lew,mi,ea}. In
this context, it is especially useful to investigate simple,
tractable models, where it is possible to relate the parameters
characterizing the nanostructure, notably the point-contact
conductances, with those that govern the magnitude of the pumped
charge. Thus by considering a small quantum dot which supports
resonant transmission, it has been shown \cite{w,lew} that when
the Fermi energy in  the leads connecting the dot with the
electronic reservoirs aligns with the quasi-bound state energy in
the dot, the charge pumped over a period is close to $e$.

In the following, the conditions for the charges pumped over a
period, $Q$, to be (almost) quantized  due to interference effects
alone will be reviewed. Our discussion begins in Sec. 2 with the
derivation of the expression for that charge. The derivation is
based on a systematic expansion of the time-dependent scattering
states around the instantaneous solution. The first-order yields
(when the system is un-biased by  constant chemical potential
differences) the widely-used `adiabatic approximation', and
reproduces the formula of Brouwer \cite{b}. The next order gives
the first correction to the adiabatic approximation. We continue
(in Sec. 3) with an analysis of the turnstile geometry. That
discussion points out to the connection between the conditions for
resonance transmission and integral values of $Q$. In Sec. 4 we
discuss the pump based on the surface acoustic waves (SAW's), and
show that interference effects lead to a staircase structure of
the acousto-current driven by the SAW's, similar to the one
observed in the experiments \cite{Sh96}. Finally, Sec. 5 includes
concluding remarks.

\section{Time-dependent Scattering Theory: Adiabatic
Expansion}

Consider a ballistic nanostructure of arbitrary geometry,
connected by an arbitrary number of leads (denoted by $\alpha$) to
massive metallic conductors which serve as  electron reservoirs
(kept at thermal equilibrium). Each reservoir is kept at a
chemical potential $\mu_{\alpha}$, such that the electrons in it
obey the Fermi distribution
\begin{eqnarray}
 f_{\alpha}(E)=\frac{1}{e^{\beta (E-\mu_{\alpha})}+1}.
\label{f}
\end{eqnarray}
The nanostructure is also subject to a potential modulated
periodically in time, $V({\bf r},t)$ which is confined to the
structure, such that asymptotic behaviors of the scattering
solutions can be defined in an un-ambiguous manner. The
Hamiltonian of the system then reads
\begin{eqnarray}
{\cal H}({\bf r},t)={\cal H}_{0}({\bf r})+V({\bf r},t),
\end{eqnarray}
in which ${\cal H}_{0}$ consists of the kinetic energy. As in the
usual scattering treatment, we seek for the scattering state
$\Psi_{\alpha n}$, which is excited by the free wave $w_{\alpha
n}^{-}$ (incoming in the transverse mode $n$ of lead $\alpha$ with
energy $E$), which is normalized to carry a unit flux,
\begin{eqnarray}
\Psi_{\alpha n}({\bf r},t)=e^{-iEt}\chi_{\alpha n}({\bf r},t)
\equiv e^{-iEt}\Bigl (w^{-}_{\alpha n}({\bf
r})+\tilde{\chi}_{\alpha n}({\bf r},t)\Bigr ). \label{scatsol}
\end{eqnarray}
By inserting this form into the time-dependent Schr\"{o}dinger
equation, (noting that $w_{\alpha n}^{-}$ is an eigenstate of
${\cal H}_{0}$), $\tilde{\chi}_{\alpha n}$ can be written in terms
of the instantaneous Green function, $G^{t}(E)$,
\begin{eqnarray}
\Bigl (E-{\cal H}({\bf r},t)\Bigr )G^{t}(E;{\bf r},{\bf
r}')=\delta ({\bf r}'-{\bf r}),\label{green}
\end{eqnarray}
as follows \cite{eay}
\begin{eqnarray}
\Bigl
(G^{t}\Bigr )^{-1}\tilde{\chi}_{\alpha n}({\bf r},t)=V({\bf
r},t)w^{-}_{\alpha n}({\bf r})-i\frac{\partial\tilde{\chi}_{\alpha
n}({\bf r},t)}{\partial t}.\label{chitild}
\end{eqnarray}
Note that the time dependence of the scattered wave function,
$\tilde{\chi}_{\alpha n}({\bf r},t)$, has the same characteristic
time scale as $V$: e.g., when the modulating potential is
oscillating in time with frequency $\omega$, $\tilde \chi$
contains all harmonics. Once the time-dependent scattering
solution is known, the thermal average of the current density
operator is given by \cite{pretre,lw,l}
\begin{equation}
\langle {\bf j}({\bf r},t)\rangle
=\frac{e}{m}{\Im}\int\frac{dE}{2\pi}\sum_{\alpha n
}f_{\alpha}(E)\chi^{\ast}_{\alpha n }({\bf
r},t)\frac{\partial\chi_{\alpha n}({\bf r},t)}{\partial {\bf r}}.
\end{equation}
Evaluating $\langle {\bf j}({\bf r},t)\rangle$ as ${\bf r}$
approaches $\infty$ in lead $\beta$, and then integrating over the
cross-section of that lead yields the current $I_{\beta}(t)$
flowing into lead $\beta$ \cite{eay,l}.

The time-dependent scattering solutions are derived from a
systematic expansion in the temporal derivatives of the
instantaneous solutions (that is, the scattering solutions of the
Hamiltonian in which time is `frozen'). Namely, Eq.
(\ref{chitild}) is solved iteratively: the temporal derivative
appearing on the right-hand-side is regarded as a small
correction. The zero-order, $\chi^{t}_{\alpha n}$, is the
scattering solution of the instantaneous Hamiltonian (in which
time appears as a parameter),
\begin{eqnarray}
\chi_{\alpha n}^{t}({\bf r})=w^{-}_{\alpha n}({\bf r})+\int d{\bf
r}'G^{t}(E;{\bf r},{\bf r}')V({\bf r}',t)w^{-}_{\alpha n}({\bf
r}').\label{chit}
\end{eqnarray}
Then the scattering solution reads
\begin{eqnarray}
\chi({\bf r},t)=\chi^{t}({\bf r}) +\chi^{(1)} ({\bf
r},t)+\chi^{(2)} ({\bf r},t)+...,\label{expansion}
\end{eqnarray}
with the first-order
\begin{eqnarray}
\chi^{(1)} ({\bf r},t)=-i\int d{\bf r}'G^{t}(E;{\bf r},{\bf
r}')\dot{\chi}^{t}({\bf r}'),\label{first}
\end{eqnarray}
and the second-order
\begin{eqnarray}
\chi^{(2)} ({\bf r},t)=-i\int d{\bf r}'G^{t}(E;{\bf r},{\bf
r}')\Delta\dot{\chi}^{t}({\bf r}'),\label{second}
\end{eqnarray}
where
\begin{eqnarray}
\Delta\dot{\chi}^{t}({\bf r}')=-i\int d{\bf r}''\frac{d}{dt}\Bigl
(G^{t}(E;{\bf r}',{\bf r}'')\dot{\chi}^{t}({\bf r}'')\Bigr ),
\end{eqnarray}
and $\dot{\chi}_{\alpha n}^{t}\equiv  d\chi_{\alpha n}^{t}/dt$.
Hence, in our iterative solution, the time-dependent scattering
states are given entirely in terms of the {\it instantaneous}
solutions of the problem at hand. This expansion procedure
necessitates that the characteristic inverse time-constant, which
describes the time dependence of the modulating potential, will be
smaller than any characteristic energy scale of the electrons.
However, it turns out that the expansion also requires that the
amplitude of the modulating potential will be small. In general,
the validity regime of the adiabatic expansion is a rather
delicate question, being related to the ratio of the time it takes
the electrons to traverse the sample (the Wigner time), and the
time-constant of the modulating potential \cite{a,ak}. In this
respect, the comparison of the results from the adiabatic
expansion, and the exact Floquet solution of the problem (when
possible), are highly desirable \cite{buttiker}.

\subsection{The Adiabatic Approximation}

Let us first analyze the net current passing through the system
utilizing the adiabatic approximation, that is, keeping only the
term (\ref{first}). For simplicity, we confine the discussion to a
nanostructure connected to left ($\ell$) and right ($r$) leads.
Then the current flowing during a single period of the modulating
potential is \cite{eay}
\begin{eqnarray}
I=\frac{1}{2}\oint \frac{dt}{T}\Bigl (I_{\ell}(t)-I_{r}(t)\Bigr
)=I_{\rm bias}+I_{\rm pump},\label{Itot}
\end{eqnarray}
where $T$ is the periodicity of the modulating potential. The
first part, $I_{\rm bias}$, flows only when the system is biased,
\begin{eqnarray}
&I_{\rm bias}=e\oint\frac{dt}{T}\int\frac{dE}{4\pi}\Bigl
(f_{\ell}(E)-f_{r}(E)\Bigr )\nonumber\\
&\times \sum_{nm}\Biggl [2|S^{t}_{rm,\ell n}|^{2} +\Re\Bigl
(S^{t}_{\ell m,\ell n}U^{\ast}_{\ell m,\ell n}-S^{t}_{\ell
m,rn}U^{\ast}_{\ell m,rn}-S^{t}_{rm,\ell n}U^{\ast}_{rm,\ell
n}+S^{t}_{rm,rn}U^{\ast}_{rm,rn}\Bigr )\Biggr ],\label{ibias}
\end{eqnarray}
where $ U_{\beta m,\alpha n}=\int d{\bf r}\chi^{t}_{\beta m}({\bf
r}) \dot{\chi}^{t}_{\alpha n}({\bf r}),$ and $S^{t}_{\beta
m,\alpha n}$ is the matrix element of the instantaneous scattering
matrix. Equation (\ref{ibias}) can be considered as a
generalization of the Landauer formula, extended to include the
effect of a time-dependent potential in the adiabatic
approximation. The second part of the current, $I_{\rm pump}$, is
established by the time-dependent potential (though it is affected
by the chemical potential difference, when the latter is applied).
Explicitly,
\begin{eqnarray}
I_{\rm pump}=e\oint\frac{dt}{T}\int\frac{dE}{4\pi} \frac{\partial
(f_{\ell}(E)+f_{r}(E))}{\partial E} \frac{1}{2}\sum_{m}\Bigl
[\langle\chi^{t}_{\ell m}|\dot{V}|\chi^{t}_{\ell m}\rangle
-\langle\chi^{t}_{r m}|\dot{V}|\chi^{t}_{r m}\rangle\Bigr ].
\label{ipump}
\end{eqnarray}
It can be shown \cite{eay} that the terms in the square brackets
of (\ref{ipump}) reproduce the Brouwer \cite{b} formula, derived
for an unbiased system (in which $I_{\rm pump}$ is given in terms
of temporal derivatives of the instantaneous scattering matrix).

\subsection{Corrections to the Adiabatic Approximation}

When the second-order in the expansion (\ref{expansion}) is
retained, one obtains the first correction to the widely-used
adiabatic approximation. We will discuss here the pumping current
beyond the adiabatic approximation for an unbiased system
connected to two single-channel leads. In that case, the current
entering lead $\beta$ consists of two parts, the leading order in
the adiabatic approximation, $I_{\beta}$, which has been discussed
above, and a correction, $\Delta I_{\beta}$. Explicitly \cite{eay}
\begin{eqnarray}
&\tilde{I}_{\beta}(t) =I_{\beta}(t)+\Delta
I_{\beta}(t),\nonumber\\
&I_{\beta}(t)=\frac{e}{2\pi}\int dE\Bigl (\frac{\partial
f(E)}{\partial E}\Bigr )\langle\chi^{t}_{\beta}|\dot{V}|\chi^{t}_{\beta}\rangle ,\nonumber\\
&\Delta I_{\beta}(t)=\Im\Bigl (\langle\chi^{t}_{\beta}|2
\dot{V}(t)\dot{G}^{t}(E)+\ddot{V}(t)G^{t}(E)|\chi^{t}_{\beta}\rangle
\Bigr ).\label{full}
\end{eqnarray}
The relative magnitude of the correction compared to the
leading-order term may be accessed by noting that \cite{eay}
\begin{eqnarray}
&\langle\chi^{t}_{\beta}|2
\dot{V}(t)\dot{G}^{t}(E)+\ddot{V}(t)G^{t}(E)|\chi^{t}_{\beta}\rangle
\nonumber\\
&=-\langle\chi^{t}_{\beta}|2\dot{V}G^{t}(E)\dot{V}+\ddot{V}|
\frac{\partial\chi^{t}_{\beta}}{\partial E}\rangle .
\end{eqnarray}
Hence, the validity of the adiabatic approximation is not only
restricted by the smallness of the inverse time-constant $T$
dominating the temporal derivatives. It depends as well on the
energy derivatives of the scattering states (i.e., the energy
scale of the instantaneous reflection and transmission amplitudes)
and the strength of the modulating potential itself \cite{wagner}.
Below, we evaluate the pumping current in the adiabatic
approximation, keeping the above restrictions in mind.

\section{Interference and Quantized Pumping in Turnstile Devices}

The correlation between resonant transmission and the magnitude of
adiabatically pumped charge may be summarized generically as
follows. Consider a quantum dot, connected to its external leads
by two point-contacts, whose conductances are controlled by
split-gate voltages which are modulated periodically in time.
During each cycle the system follows a closed curve, the `pumping
contour', in the parameter plane spanned by the point-contact
conductances. As the system parameters are varied (for example,
the gate voltage on the dot) the pumping contour distorts and
shifts, forming a Lissajous curve in the parameter plane. The
pumped charge will be (almost) quantized when the pumping contour
encircles transmission peak(s) (that is, resonances) of the
quantum dot in that parameter plane. Its magnitude (in units of
the electronic charge, $e$) and sign are determined by the winding
number of the pumping contour.

In order to exemplify this idea, we employ the tight-binding
description, and imagine the quantum dot to be coupled to
semi-infinite, one-dimensional and single-channel leads by matrix
elements $J_{\ell}$ and $J_{r}$. Those are oscillating in time
with frequency $\omega$, such that the modulation amplitude is P
and the phase shift between the $J_{\ell}$ modulation and that of
$J_{r}$ is $2\phi$,
\begin{eqnarray}
J_{\ell}=J_{\rm L}+{\rm P}\cos (\omega t +\phi ), \ \  J_{\rm
r}=J_{\rm L}+{\rm P}\cos (\omega t -\phi ) .\label{jljr}
\end{eqnarray}
The point-contact conductances are then given (in dimensionless
units) by X$_{\ell}\equiv J_{\ell}^{2}$ and X$_{r}\equiv
J_{r}^{2}$. As the couplings of the quantum dot to the leads are
modulated in time, $J_{\ell}$ and $J_{r}$ can attain both negative
and positive values. This reflects a modulation of the potential
shaping the dot: The tight-binding parameters $J_\ell$ and $J_{\rm
r}$, which are derived as integrals over the site `atomic' wave
functions and the oscillating potential, can have both signs. The
extreme modulation arises for $J_{\rm L}=0$, when the hopping
matrix elements which couple the dot to the leads oscillate in the
range $\{$-P,P$\}$. The conductances of the point contacts are
then modulated in the range $\{$0,P$^{2}\}$. The corresponding
Lissajous curve of the pumping is then a simple closed curve.
Another possibility is to keep the couplings finite at any time,
i.e., $J_{\rm L}\neq 0$, and to modulate the couplings around this
value. In that case the Lissajous curve may fold on itself
[compare Figs. 2 and 3 below]. This yields a rather rich behavior
of the pumped charge.

Indeed, using the expression for the charge, Q, pumped through the
quantum dot during a single period of the modulation
(\ref{ipump}),
\begin{eqnarray}
{\rm Q}=\frac{e}{4\pi}\oint dt\Bigl
[\langle\chi^{t}_{r}|\dot{V}|\chi^{t}_{r}\rangle
-\langle\chi^{t}_{\ell}|\dot{V}|\chi^{t}_{\ell}\rangle\Bigr ],
\label{charge}
\end{eqnarray}
(the limit of zero temperature is taken for simplicity) where
$\chi^{t}_{\ell ,r}$ are the instantaneous scattering solutions
incited by free waves from the left ($\ell$) and from the right
($r$), the variation of $Q$ as function of $P$ is, e.g.,

\begin{figure}[htb]
\begin{center}
\includegraphics[scale=0.75]{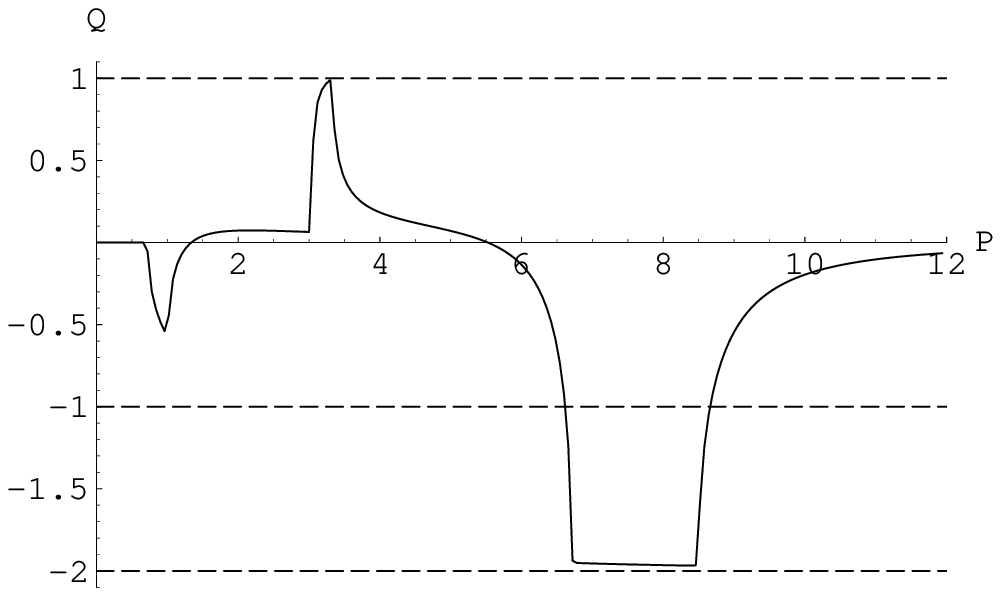}\ \ \  \ \includegraphics[scale=0.75]{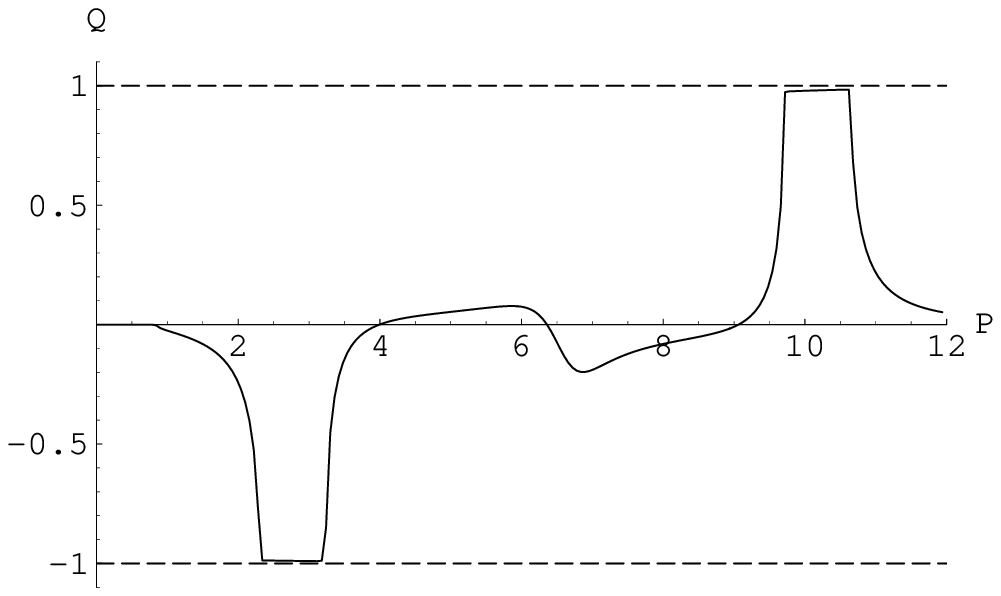}
\end{center}
\caption{The pumped charge, Q, in units of $e$, as function of the
modulating amplitude, P.} \label{fig1}
\end{figure}

\noindent(In drawing these figures, we have used a tight-binding
description for the entire system, and varied the model
parameters, see Ref. \cite{ea}.)

\begin{figure}[htb]
\begin{center}
\includegraphics[scale=0.3]{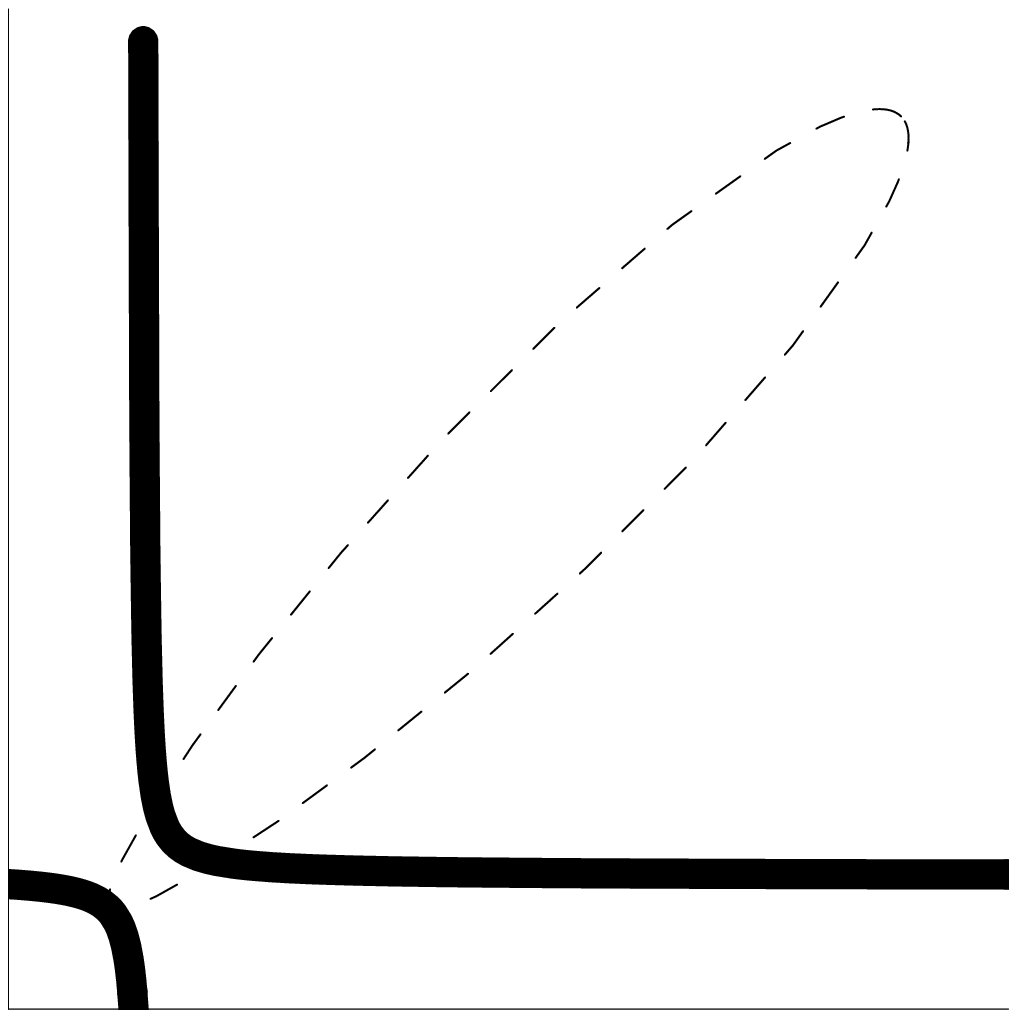}\ \ \  \ \includegraphics[scale=0.3]{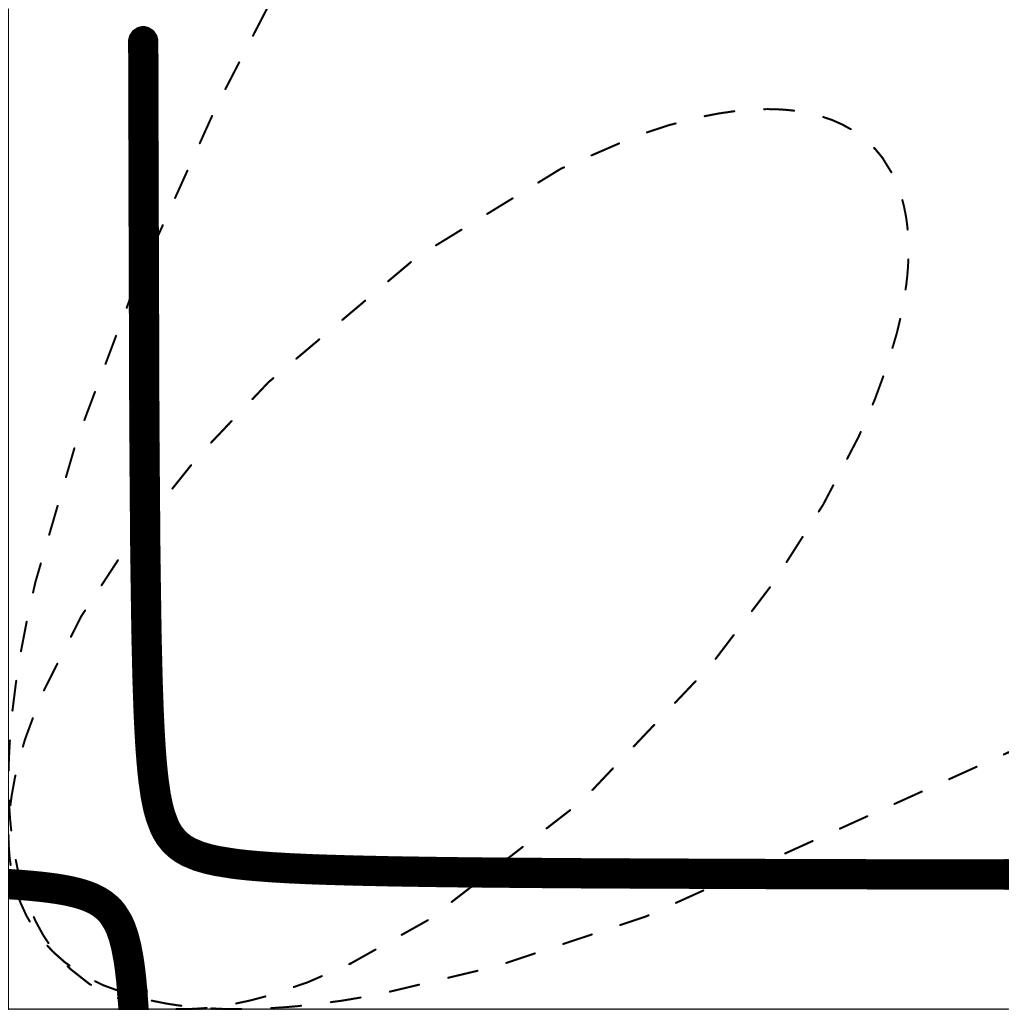}
\ \  \ \includegraphics[scale=0.3]{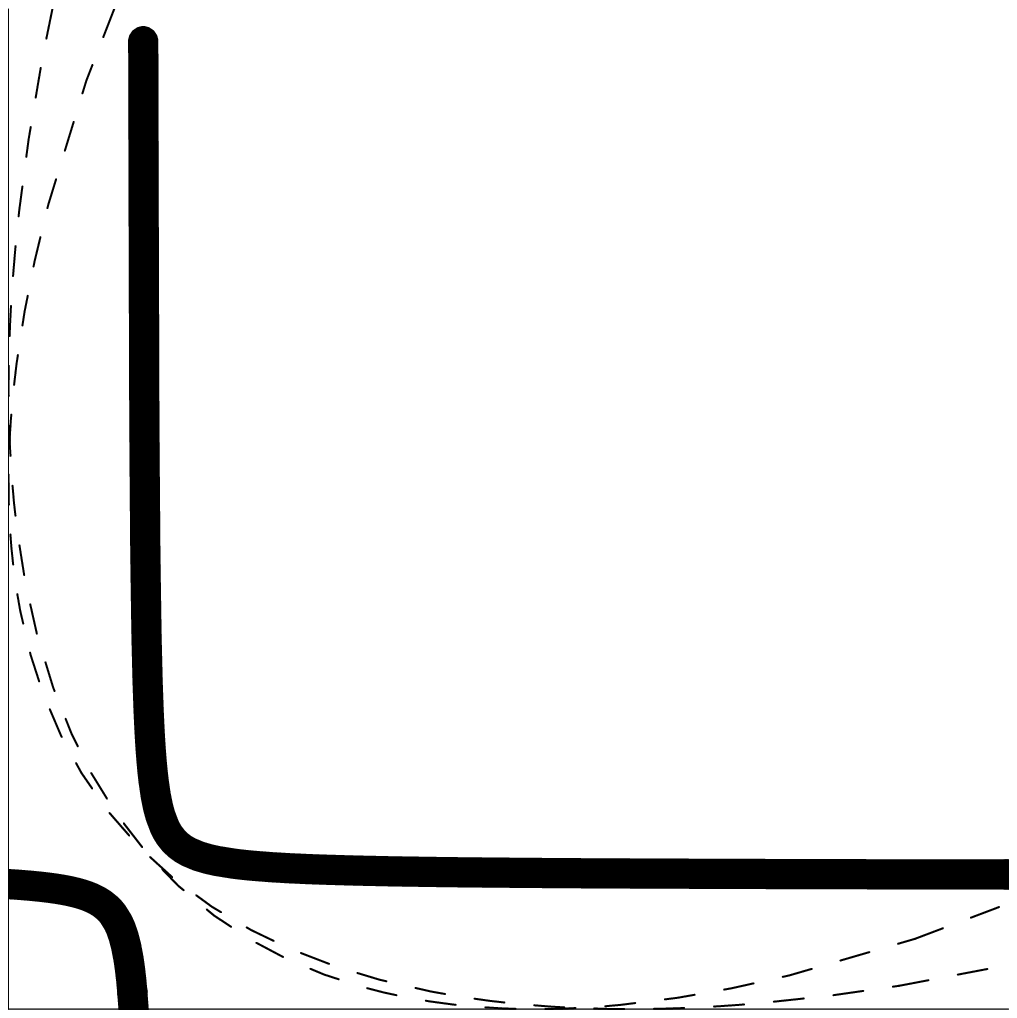}
\end{center}
\caption{The pumping contour (dashed line) and the resonance lines
(thick lines), in the \{X$_{\ell}$-X$_{r}$\} plane, for P=1, (left
panel), P=5 (central panel) and P=8 (right panel). } \label{fig2}
\end{figure}

In order to relate the `quantized' values of $Q$, as portrayed for
example on the left panel of  Fig. \ref{fig1}, and the location of
the pumping contour relative to the resonant transmission of the
quantum dot, we exhibit in Figs. \ref{fig2} the pumping contour in
the parameter plane, and the resonances lines, for several values
of $P$.

The resonance lines are found by calculating the transmission
through the quantum dot as function of X$_{\ell}$ and X$_{r}$. One
then finds \cite{ea} that when either X$_\ell$ or X$_r$ is varied
while the other parameter is kept fixed, the maxima of the
transmission occur on two `resonance lines', shown in thick lines
in Figs. \ref{fig2} . The figures also show the topology of the
curve traversed by the system during the pumping cycle for
representative values of P. When P=1, the pumping contour encloses
a small part of the upper resonance line, and also touches the
peak on the lower resonance line. Indeed, Q has an intermediate
value near $-0.5$, decreasing to zero as P moves away from 1 (see
Fig. \ref{fig1}, left panel). Increasing the amplitude to the
value P=5 reveals that the pumping contour encloses both peaks on
the resonance lines, and therefore their separate contributions
almost cancel one another, leading to a tiny value of Q. Also, the
pumping curve begins to fold on itself, giving rise to the
`bubble' close to the origin (see central and left panels in Fig.
\ref{fig2}). Following the increase of that bubble as P is
enhanced leads to the situation in which the bubble encloses the
lower resonance, and then the charge attains a unit value (see
left panel on Fig. \ref{fig1}). As the bubble increases further,
capturing the two resonance lines, Q again becomes very small. But
upon further increasing P, we reach the interesting situation,
depicted in Fig. \ref{fig2} for P=8, in which the bubble encircles
{\it twice} the upper resonance line, leading to a pumped charge
very close to $|2e|$ (see left panel on Fig. \ref{fig1}).

Thus, the condition for obtaining integral values of the pumped
charge is that the contour traversed by the system in the
parameter plane spanned by the pumping parameters should encircle
a significant portion of a resonance line in that plane. The
magnitude and the sign of the pumped charge are determined by that
portion, and by the direction along which the resonance line is
encompassed. We emphasize that the pumping contour, as well as the
resonance lines, can be determined experimentally \cite{lew}.

The reason for this topological description of adiabatic charge
pumping can be traced back to the expression for the pumped
charge, Eq. (\ref{charge}). The main contribution to the temporal
integration there comes from the poles of the integrand. The same
poles are also responsible for the resonant states of the
nanostructure, that is, for the maxima in the transmission
coefficient \cite{lew,ea}.  One can imagine more complex
scenarios: Including higher harmonics of $\omega$ in the time
dependence of the point contact conductances can create more
complex Lissajous contours, which might encircle portions of the
resonance lines more times, yielding higher quantized values of
the pumped charge.

Finally, it should be mentioned that the results presented above
are obtained at zero temperature. At finite temperatures, the
expression for Q should be integrated over the the electron energy
$E$, with the Fermi function derivative $-\partial f/\partial E$.
Hence, upon the increase of the temperature, the pumped charge
would be smeared and suppressed. It would be very interesting to
check the above predictions in more complicated models (for
example, when there are several levels on each of the sites
forming the quantum dot), allowing for a richer structure of the
transmission in the parameter space, and, of course, in real
systems.

\section{Pumping by Surface Acoustic Waves in a Quantum Channel}

Modulation of the potential acting on a nanostructure may be also
achieved through the piezoelectric effect of surface acoustic
waves (SAW's), which is relatively large in GaAs. In the
two-dimensional electron gas formed in GaAs-AlGaAs samples the
potential created by the SAW is screened out. However, the SAW's
are effective within a quasi-one-dimensional channel (where
screening is diminished) defined in GaAs-AlGaAs samples. In the
experiments \cite{Sh96}, the time-averaged current exhibits steps
between plateaus, at quantized values of integer$\times e(\omega
/2\pi )$, ($\omega $ is the SAW frequency), as function of either
the gate voltage on the quantum channel, or the SAW amplitude.
Here we propose an explanation for this observation, in terms of
interference of non-interacting electrons \cite{SAW1,SAW2}.

Unlike the turnstile-like case, the piezoelectric potential,
${\cal H}_{\rm SAW}({\bf r},t)=$P$ \cos(\omega t-{\bf q}\cdot{\bf
r})$, generated by the SAW oscillates with time everywhere inside
the nanostructure. The induced average current (in the absence of
bias), flows in the direction of the SAW wave vector, ${\bf q}$. A
realistic treatment of the experimental geometry \cite{SAW2} only
allowed a calculation at low SAW amplitude, P, yielding $Q \propto
$P$^2$. The screening of the piezoelectric potential in the wide
banks of the channel is also difficult to treat exactly. In view
of this, we have proposed a simple model \cite{SAW1}, in which the
quantum channel is described by a finite one-dimensional
tight-binding chain, whose on-site energies are modulated in time
\begin{equation}
\epsilon_n(t)=V+ {\rm P} \cos[\omega t- qa(n-n_0)]. \label{eps}
\end{equation}
(Effects due to gradual screening, or reflections from the channel
ends, can be incorporated as well \cite{SAW1}.) Here $V$
represents the gate voltage and P$>0$, so that $\epsilon_n$ has a
maximum (minimum) in the center of the channel $n_0=(N+1)/2$ at
$t=0$ ($T/2$). Within this simplified model, the charge pumped
within each period of the modulating potential can be written as
\cite{SAW1}
\begin{equation}
Q=\frac{e J_{\rm L}^2 \sin ka}{\pi J} \int_{-T/2}^{T/2} dt
\sum_{n=1}^N \dot{\epsilon}_{n}|g_{n,1}|^{2},
\label{int}
\end{equation}
where $J_{\rm L}$ denotes the coupling of the finite chain to the
leads, $J$ is the tight-binding coupling on the leads, $k$ is the
wave vector of the incoming scattering state (corresponding to the
Fermi energy), and $g_{\ell ,m}$ is the instantaneous Green
function matrix of the finite chain, (with the coupling to the
infinite leads included as a self-energy).

\begin{figure}[htb]
\begin{center}
\includegraphics[scale=0.7]{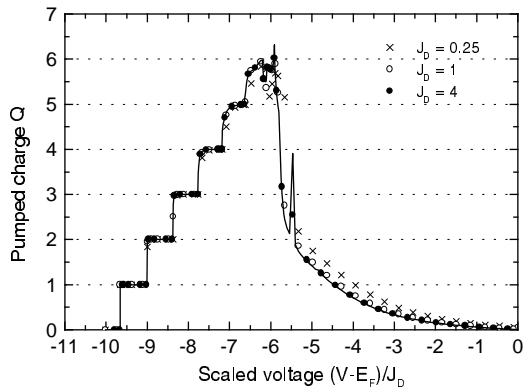}
\end{center}
\caption{The pumped charge {\it vs.} the gate voltage, for a SAW
wave length equal to 4 times the channel length. } \label{fig3}
\end{figure}

As has been found for the turnstile device in the previous
section, the charge pumped by the SAW correlates with the behavior
of the time-average transmission through the one-dimensional
channel \cite{SAW1}: Wherever that transmission shows spikes, as
function of the tight-binding parameters, $Q$ exhibits large
changes. The typical variation of $Q/e$, as function of the gate
voltage $V$, is depicted in Fig. \ref{fig3}, demonstrating that a
staircase structure, reminiscent of the one observed in the
experiments \cite{Sh96}, can  result from interference effects
alone.

\begin{figure}[htb]
\begin{center}
\includegraphics[scale=0.7]{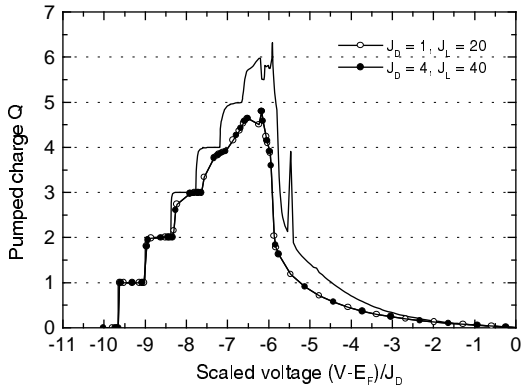}
\end{center}
\caption{The deterioration of the staircase structure upon
increasing the coupling to the leads (scaled by the tight-binding
parameter J$_{\rm L}$). } \label{fig4}
\end{figure}

Our simplified model allows for a detailed study of the robustness
of the staircase structure against variation of the couplings to
the leads, of the SAW amplitude P, of the Fermi energy, and other
parameters. As an example, we portray in Fig. \ref{fig4} the
deterioration of the staircase structure upon increasing the
coupling to the leads. That increase amounts to a broadening of
the resonant levels within the channel, and consequently to the
reduction of the quantization.

\begin{figure}[htb]
  \begin{center}
  \includegraphics[scale=0.7]{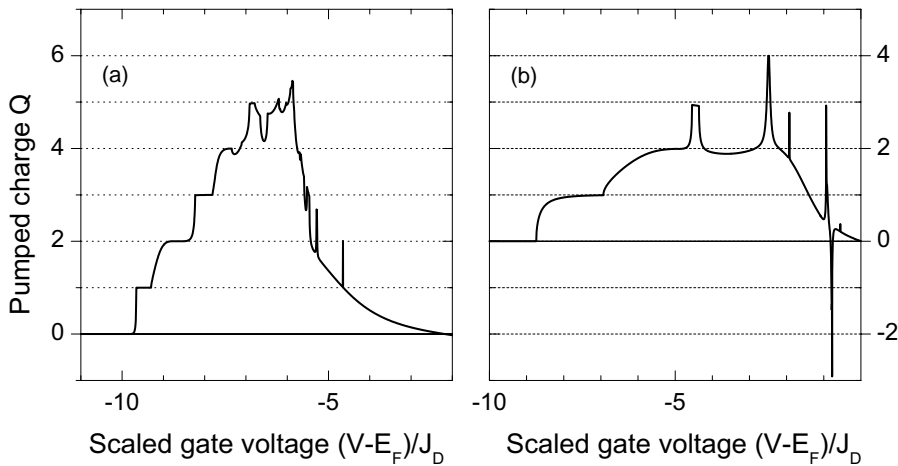}
  \end{center}
  \caption{
  (a) The effect of a static disorder on the staircase structure.
  Random on-site energies $V_n$, drawn from a uniform
  distribution $[ -J_d, J_d ]$ have been added to $\epsilon_n$.
  (b) Effects of SAW damping within the channel, modelled
  by multiplying the second term in Eq.~(\ref{eps}) with a Gaussian profile,
  $e^{-[a(n-n_0)/l_0]^2}$,  $l_0=L/4$.} \label{fig5}
\end{figure}

The  quantization is also stable with respect to a random noise in
the channel, or a gradual screening of the SAW, as shown in Fig.
\ref{fig5}. The shape  and the interval between the steps may vary
as the pumping potential is modified, but the $Q(V)$ curve follows
the same generic pattern as a function of the SAW amplitude $P$:
the first step appears once a certain threshold value of $P$ is
exceeded, and the number of steps increases gradually as $P$ is
further elevated above the threshold.

\begin{figure}[htb]
\begin{center}
\includegraphics[scale=0.7]{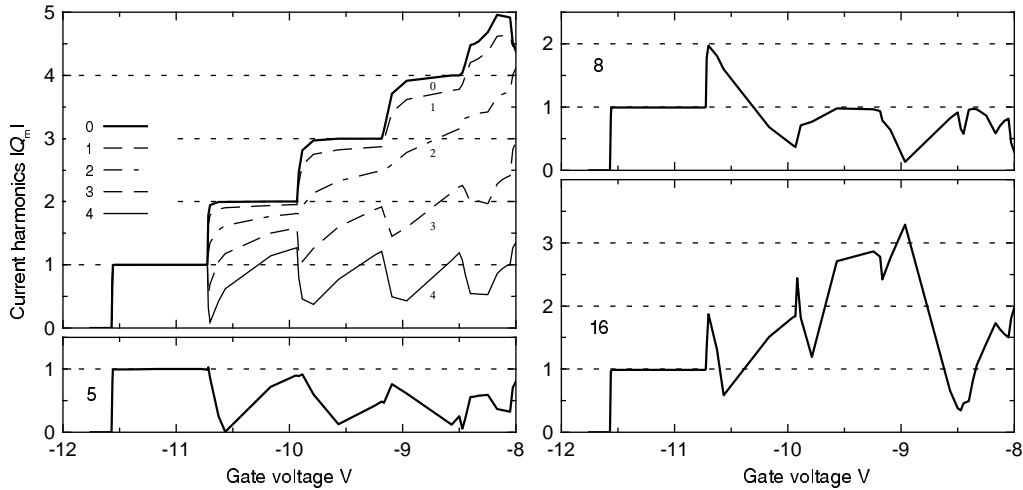}
\end{center}
\caption{Higher harmonics of the pumped charge.} \label{fig6}
\end{figure}

Another interesting feature is that when the quantization is due
to interference (as opposed to the Coulomb-blockade effect), the
staircase structure appears also in the amplitude of the higher
harmonics, as exemplified in Fig. \ref{fig6}.

To conclude this section, we note that even within this simple 1D
model, it is possible to study effects arising from variations of
the SAW amplitude $P$ in space (due to screening effects, or to
multiple reflections from the channel's ends), or from random
energies $\{V_n\}$, which may represent impurities within the
channel.

\section{Concluding remarks}

Using simplified models, we have demonstrated that interference
effects suffice to produce transfer of integral number of
electrons through a mesoscopic (unbiased) system, subject to a
periodically-varying potential. The calculations presented above
utilize the expression for the pumped charge, $Q$, in the
adiabatic approximation, that is, keeping only the first-order of
the expansion (\ref{expansion}). However, the next-order
corrections need not be small; It is therefore an open question
whether the quantization, and its relation to resonant
transmission, will still be present when higher terms in the
temporal expansion are retained. We hope to pursue these issues in
the future.

\section*{Acknowledgements}
We thank Y. Imry, Y. Levinson, and P. W\"{o}lfle for helpful
conversations. This research was carried out in a center of
excellence supported by the Israel Science Foundation, and was
supported in part  by the Albert Einstein Minerva Center for
Theoretical Physics at the Weizmann Institute of Science.

\end{document}